\documentclass[hyper]{JHEP} 

\usepackage{epsfig}

\newcommand\fverb{\setbox\pippobox=\hbox\bgroup\verb}

\newcommand\fverbdo{\egroup\medskip\noindent%

            \fbox{\unhbox\pippobox}\ }

\newcommand\fverbit{\egroup\item[\fbox{\unhbox\pippobox}]}

\newbox\pippobox

\title{Note About Hamiltonian
Structure of  Non-Linear Massive
Gravity}
\author{J. Kluso\v{n}\\
Department of
Theoretical Physics and Astrophysics\\
Faculty of Science, Masaryk University\\
Kotl\'{a}\v{r}sk\'{a} 2, 611 37, Brno\\
Czech Republic\\
E-mail: \email{klu@physics.muni.cz}}
\preprint{}

 \abstract{We perform the Hamiltonian analysis of non-linear massive
 gravity action studied recently
 in arXiv:1106.3344 [hep-th].
We show that  the Hamiltonian
constraint is the second class
constraint. As a result the theory
possesses an odd number of the second
class constraints and hence all non
physical degrees of freedom cannot be
eliminated.} \keywords{ Massive
Gravity, Hamiltonian Formalism }

\def\bA{\mathbf{A}}
\def\bB{\mathbf{B}}

\def\tmH{\tilde{\mH}}
\def\be{\begin{equation}}
\def\bD{\mathbf{D}}
\def\ee{\end{equation}}
\def\bD{\mathbf{D}}
\def\bea{\begin{eqnarray}}

\def\eea{\end{eqnarray}}

\def\tr{\mathrm{tr}\, }

\def\bI{\mathbf{I}}

\def\mH{\mathcal{H}}

\def\tr{\mathrm{Tr}}

\def\bx{\mathbf{x}}
\def\by{\mathbf{y}}

\newcommand{\hg}{\hat{g}}

\newcommand{\mG}{\mathcal{G}}

\def \bA{\mathbf{A}}

\newcommand{\bT}{\mathbf{T}}

\newcommand{\mL}{\mathcal{L}}

\def\pb #1{\left\{#1\right\}}

\begin{document}
\section{Introduction}\label{first}
One of the most challenging problem is
to find consistent formulation of
massive gravity. The first attempt for
construction of  this theory is dated
to the year 1939 when Fierz and Pauli
formulated its version of linear
massive gravity \cite{Fierz:1939ix}
\footnote{For review, see
\cite{Hinterbichler:2011tt}.}. However
it is very non-trivial task to find  a
consistent non-linear generalization of
given theory and it remains as an
intriguing theoretical problem. It is
also important to stress that recent
discovery of dark energy and associated
cosmological constant problem has
prompted investigations in the long
distance modifications of general
relativity, for review, see
\cite{Clifton:2011jh}.

Returning to the theories of massive
gravity we should mention that these
theories suffer from the problem of the
ghost instability, for very nice
review, see \cite{Rubakov:2008nh}.
Since the general relativity is
completely constrained system there are
four constraint equations along the
four general coordinate transformations
that enable to eliminate four of the
six propagating modes of the metric,
where the propagating mode corresponds
to a pair of conjugate variables.  As a
result the number of physical degrees
of freedom is equal to two which
corresponds to the massless graviton
degrees of freedom. On the other hand
in case of the massive gravity the
diffeomorphism invariance is lost and
hence the theory contains six
propagating degrees of freedom which
only five correspond to the physical
polarizations of the massive graviton
while the additional mode is ghost.

It is natural to ask the question
whether it is possible to construct
theory of massive gravity where one of
the constraint equation and associated
secondary constraint eliminates the
propagating scalar mode. It is
remarkable that linear Fierz-Pauli
theory does not suffer from the
presence of such a ghost. On the other
hand it was shown by Boulware and Deser
\cite{Boulware:1973my} that ghosts
generically reappears at  the
non-linear level. However it was shown
recently by de Rham and Gobadadze in
\cite{deRham:2010ik} that it is
possible to find such a formulation of
the massive gravity which is a ghost
free in the decoupling limit. Then it
was also shown in \cite{deRham:2010kj}
that this action that was written in
the perturbative form can be ressummed
into fully non-linear actions. It was
claimed there that this is the first
successful construction of potentially
ghost free non-linear actions of
massive gravity.

However it is still an open problem
whether this theory contains ghost or
not, for discussion, see for example
\cite{Chamseddine:2011mu}. On the other
hand S.F. Hassan and R.A. Rosen showed
recently in \cite{Hassan:2011hr} on the
non-perturbative level that it is
possible to perform such a redefinition
of the shift function so that the
resulting theory still contains the
Hamiltonian constraint. Then it was
argued that the presence of this
constraint allows to eliminate the
scalar mode and hence the resulting
theory is the ghost free massive
gravity \footnote{ For another works
that support the claim that the
non-linear gravity action is ghost
free, see
\cite{D'Amico:2011jj,deRham:2011qq}.}.

In this paper we again perform the
Hamiltonian analysis of the non-linear
massive gravity  action presented in
\cite{Hassan:2011vm}. The important
point which was not addressed in this
paper  is the character of the
Hamiltonian constraint. In fact, the
scalar mode can be eliminated on
condition that the Poisson brackets
between Hamiltonian constraint and all
constrains vanish on the constraint
surface since then either the time
evolution of the Hamiltonian constraint
is trivially zero or it induces an
additional constraint. The first case
corresponds to the situation when the
Hamiltonian constraint is the first
class constraint while the second case
corresponds situation when the
Hamiltonian constraint together with
additional constraint are the second
class constraints.
 However we show
that the Poisson brackets between
Hamiltonian constraints and also
between Hamiltonian constraints and
some of other constraints  do not
vanish at generic points of the
constraint surface. In other words the
requirement of the preservation of the
Hamiltonian constraint fixes
corresponding Lagrange multipliers and
no new constraints are  generated.
We should stress that the similar
result was derived recently in case of
the Hamiltonian analysis of
non-projectable version of
Ho\v{r}ava-Lifshitz gravity
\cite{Horava:2009uw}, for very nice
discussion,  see for example
\cite{Henneaux:2010vx,Henneaux:2009zb,Li:2009bg}.
In fact, it was shown there that the
Hamiltonian constraint of
non-projectable Ho\v{r}ava-Lifshitz
gravity is the second class constraint
at generic points of the phase space
and also that the Hamiltonian strongly
vanishes which makes the physical
meaning of non-projectable version of
Ho\v{r}ava-Lifshitz gravity un clear.
 In case of the
massive gravity the situation is
slightly different since the
Hamiltonian is not given as the linear
combination of constraints and hence
does not vanish on the constraint
surface. However the fact that the
Hamiltonian constraint is the second
class means that it is not possible to
eliminate all additional physical mode.
Moreover, it is not completely clear
how to physically interpret the
additional $1/2$ degree of freedom in
the phase space.
The structure of this paper is as
follows. In the next section
(\ref{second}) we consider the action
for the general relativity with the
additional term that gives  the mass
for the graviton when we analyze the
perturbative spectrum around the flat
space-time. Then we perform the
Hamiltonian analysis of given theory
and we show that it  possesses eight
second class constraints. This result
is the manifestation of the fact that
the diffeomorphism invariance is
completely  broken which implies that
there are no first class constraints.
In section
 (\ref{third}) we perform the
 Hamiltonian analysis of the non-linear
 massive gravity with redefined shift
 functions. We show that the momentum
 conjugate to the lapse function is
 still the first class constraint
 however the Hamiltonian constraint is
 the second class constraint. This
 result implies that the physical phase
 space is odd dimensional.
  In conclusion (\ref{fourth}) we outline
 our result and suggest possible
 extension of this work.
  Finally in the Appendix
 (\ref{appendix}) we give an example of
the system with the single second class
constraint.
\section{Massive Gravity}\label{second}
We begin our analysis with the
introduction of the notation used in
\cite{Hassan:2011hr,Hassan:2011vm}. It
is well known that the Fierz-Pauli
theory, which is linearized general
relativity in flat space is extended by
the additional mass term for the metric
fluctuations $h_{\mu\nu}=
\hg_{\mu\nu}-\eta_{\mu\nu}$
\begin{equation}
\frac{m^2}{4}(h_{\mu\nu}h^{\mu\nu}-h^\mu_\mu
h^\mu_\nu) \ ,
\end{equation}
where $\mu,\nu=0,\dots,3$ and we use
the convention where the flat Minkowski
metric
$\eta_{\mu\nu}=\mathrm{diag}(-1,1,1,1)$.
In order to construct non-linear
generalization of Fierz-Pauli theory an
additional extra rank two tensor
$f_{\mu\nu}$ is introduced.
Then the
general form of the massive gravity
action is \cite{Hassan:2011vm}
\begin{equation}\label{Smassive}
S=M_p^2\int d^4x \sqrt{-\hg}R(\hg)-
M_p^2 m^2\int d^4x
\sqrt{-\hg}F(\hg^{-1}f) \ .
\end{equation}
Note that by definition $\hg^{\mu\nu}$
and $f_{\mu\nu}$ transform under
general diffeomorphism transformations
$x'^\mu=x'^\mu(x)$ as
\begin{equation}
\hg'^{\mu\nu}(x')= \hg^{\rho\sigma}(x)
\frac{\partial x'^\mu}{\partial
x^\sigma} \frac{\partial
x'^\nu}{\partial x^\sigma} \ , \quad
f'_{\mu\nu}(x')=
f_{\rho\sigma}(x)\frac{\partial
x^\rho}{\partial x'^\mu}\frac{\partial
x^\sigma}{\partial x'^\nu} \  .
\end{equation}
It  is convenient to parameterize the
tensor $f_{AB}$ using four scalar
fields $\phi^A$ and some fixed
auxiliary metric
$\bar{f}_{\mu\nu}(\phi)$ so that
\begin{equation}
f_{\mu\nu}=\partial_\mu\phi^A\partial_\nu
\phi^B \bar{f}_{AB}(\phi) \ ,
\end{equation}
where the metric $f_{AB}$ is invariant
under diffeomorphism transformation
$x'^\mu=x^\mu(x')$ which however
transforms as a tensor under
$\phi'^A=\phi'^A(\phi^B)$.
The special case $f_{AB}=\eta_{AB}$
corresponds to recently developed Higgs
gravity
\cite{'tHooft:2007bf,Chamseddine:2011mu,Alberte:2010it,Chamseddine:2010ub,Iglesias:2011it}
\footnote{For Hamiltonian analysis of
given theory, see
\cite{Kluson:2010qf}.}. In this note we
instead consider the unitary gauge
fixing of given theory when
\begin{equation}\label{gaugefixing}
\phi^A=x^\mu\delta_\mu^A \ , \quad
f_{AB}=\eta_{AB} \ .
\end{equation}
The gauge fixing (\ref{gaugefixing})
implies  that the action
(\ref{Smassive})
 is not diffeomorphism invariant and
 hence  there is no gauge
freedom left. Finally, the scalar
function $F$
 that is generally non-linear
function of the metric components
 gives the mass for the
graviton when we analyze the
fluctuations around the flat
space-time.

 In order
to find the Hamiltonian formulation of
given theory we consider ADM
formulation of gravity
\cite{Arnowitt:1962hi}, for review, see
\cite{Gourgoulhon:2007ue}.
 Explicitly,
we use  $3+1$  decomposition of the
four dimensional metric components
\begin{eqnarray}
\hat{g}_{00}&=&-N^2+N_i g^{ij}N_j \ ,
\quad \hat{g}_{0i}=N_i \ , \quad
\hat{g}_{ij}=g_{ij} \ ,
\nonumber \\
\hat{g}^{00}&=&-\frac{1}{N^2} \ , \quad
\hat{g}^{0i}=\frac{N^i}{N^2} \ , \quad
\hat{g}^{ij}=g^{ij}-\frac{N^i N^j}{N^2}
\ ,
\nonumber \\
\end{eqnarray}
where $i,j,k,\dots,=1,2,3$. Note also
that $4-$dimensional scalar curvature
has following decomposition
\begin{equation}\label{Rdecom}
{}^{(4)}R=K_{ij}\mG^{ijkl}K_{kl}+{}^{(3)}R
\ ,
\end{equation}
where ${}^{(3)}R$ is three-dimensional
spatial curvature, $K_{ij}$ is
extrinsic curvature defined as
\begin{equation}
K_{ij}=\frac{1}{2N} (\partial_t g_{ij}
-\nabla_i N_j-\nabla_j N_i) \ ,
\end{equation}
where $\nabla_i$ is covariant
derivative built from the metric
components $g_{ij}$ and where
$\mG^{ijkl}$ is De Witt metric
\begin{equation}
\mG^{ijkl}=\frac{1}{2}(g^{ik}g^{jl}+
g^{il}g^{jk})-g^{ij}g^{kl} \ .
\end{equation}
 Finally note that
 we omitted terms
proportional to the covariant
derivatives in (\ref{Rdecom}). These
terms  induce the boundary terms that
vanish for suitable chosen boundary
conditions.
Now for the case when $f_{\mu\nu}=
\mathrm{diag}(-1,1,1,1)$
 the matrix $(\hg f)^\mu_\nu$
takes the form
\begin{eqnarray}\label{mathgf}
(\hg f)^\mu_\nu=
\left(\begin{array}{cc}
\frac{1}{N^2} &
\frac{N^i}{N^2}\delta_{ij}
 \\
-\frac{N^i}{N^2} &
 (g^{ik}-\frac{N^i
N^k}{N^2})\delta_{kj} \\
 \end{array}\right) \ .  \nonumber \\
\end{eqnarray}
The presence of this term in the action
manifestly breaks the diffeomorphism
invariance of given theory. In fact,
the action now takes the form
\begin{eqnarray}\label{act2}
S=M_p^2\int d^4x \sqrt{g}N
[K_{ij}\mG^{ijkl}K_{kl}+{}^{(3)}R -m^2
F(N,N^i,g_{ij})] \ .
\end{eqnarray}
From  (\ref{act2}) it is
straightforward to find corresponding
Hamiltonian
\begin{equation}
H=\int d^3\bx (N\mH_T+N^i \mH_i) \ ,
\end{equation}
where
\begin{eqnarray}
\mH_T&=&\frac{1}{\sqrt{g}M_p^2}
\pi^{ij}\mG_{ijkl}\pi^{kl}-M_p^2
\sqrt{g} {}^{(3)}R+\nonumber
\\
&+& M_p^2 m^2\sqrt{g} F(N,N^i,g_{ij}) \ , \nonumber \\
\mH_i&=&-2g_{ik}\nabla_j\pi^{jk} \ .  \nonumber \\
\end{eqnarray}
Due to the presence of the mass term
this action is highly non-linear in
$N,N^i$  which would imply the absence
of the first class constraints.
Explicitly, since the action
(\ref{act2}) does not contain the time
derivatives of $N,N^i$ we obtain that
there are  following primary
constraints
\begin{equation}\label{primcon1}
\pi_N\approx 0 \ , \quad \pi_i\approx 0
\ .
\end{equation}
Note that these momenta have following
non-zero Poisson brackets with
conjugate coordinates
\begin{eqnarray}\label{pbNpN}
\pb{N(\bx),p_N(\by)}&=&\delta(\bx-\by)
\ , \quad \pb{N^i(\bx),\pi_j(\by)}=
\delta^i_j\delta(\bx-\by)  \ . \nonumber \\
\end{eqnarray}
As the next step we have to check the
preservation of the primary constraints
(\ref{primcon1}) during the time
evolution of the system. Using
(\ref{pbNpN}) we obtain
\begin{eqnarray}
\partial_t\pi_N&=&\pb{\pi_N,H}=
-\mH_T-NM_p^2m^2\sqrt{g}\frac{\delta
F}{\delta N}\equiv -
\tilde{\mH}_T\approx 0 \ ,
\nonumber \\
\partial_t \pi_i&=&\pb{\pi_i,H}=
-\mH_i- NM_p^2m^2\sqrt{g}\frac{\delta
F}{\delta N^i}\equiv
-\tilde{\mH}_i\approx 0 \ .
 \nonumber \\
\end{eqnarray}
The crucial point of the massive
gravity for the general form of the
function $F$ is that the secondary
constraints $\tilde{\mH}_T\approx 0,
\tilde{\mH}_i\approx 0 $ together with
the primary constraints $\pi_N\approx 0
,\pi_i\approx 0 $ are the second class
constraints due to the fact that they
depend non-trivially on $N, N^i$.
Explicitly, we have
\begin{eqnarray}
\pb{\pi_N,\tilde{\mH}_T}&=& -M_p^2
m^2\sqrt{g}N\frac{\delta^2
F}{\delta^2N} \neq 0 \ , \nonumber \\
 \pb{\pi_i,\tilde{\mH}_T}&=& -M^2_p m^2
\sqrt{g}N\frac{\delta F}{\delta
N^i}+NM_p^2 m^2\sqrt{g}\frac{\delta^2
F}{\delta N \delta N^i}\neq 0 \ ,
\nonumber \\
 \pb{\pi_N,\tilde{\mH}_i}
&=&-M_p^2 m^2\sqrt{g}\frac{\delta
F}{\delta N^i}+NM_p^2 m^2\sqrt{g}
\frac{\delta^2 F}{\delta N \delta
N^i}\neq 0 \ , \nonumber \\
\pb{\pi_i,\tilde{\mH}_j}&=& -NM_p^2
m^2\sqrt{g}\frac{\delta^2 F}{\delta N^i
\delta N^j}\neq 0 \ .  \nonumber \\
\end{eqnarray}
As a result we have following
collection of the second class
constraints
$\pi_N,\pi_i,\tilde{\mH}_T,\tilde{\mH}_i$.
The fact that $\pi_N,\pi_i$ are the
second class constraints means  that
the conjugate momenta $\pi_N,\pi_i$
vanish strongly. Then  we solve the
constraints
$\tilde{\mH}_T=0,\tilde{\mH}_i=0$ for
$N,N^i$ that could be expressed
 as  functions of canonical
variables $g_{ij},\pi^{ij}$, at least
in principle. In other words the
dynamical content of theory is given by
$6$ modes and their conjugate momenta.
On the other hand the massive graviton
has $5$-degrees of freedom so that the
additional mode is the well known
scalar mode with possible pathological
behavior.

\section{Hamiltonian Analysis of
Non-Linear  Massive Gravity}
 \label{third}
It was argued in \cite{Hassan:2011hr}
that in the specific model of the
massive gravity proposed in
\cite{Hassan:2011vm,deRham:2010kj} it
is possible to perform suitable
redefinition of the shift function
$N^i$ in such a way so that the
resulting theory of massive gravity
possesses additional constraints. Then
it was argued there that the  presence
of these
 constraints eliminates the additional
scalar mode leaving the physical
spectrum of massive gravity only.

Our goal is to perform an explicit
Hamiltonian analysis of this theory
with redefined shift functions in order
to determine its  constraint structure.
For our purposes it is sufficient to
consider following simple model of the
non-linear massive gravity action
\cite{Hassan:2011vm,deRham:2010kj}
\begin{eqnarray}\label{nonmassact}
S=M_p^2\int d^4x \sqrt{g}N
[K_{ij}\mG^{ijkl}K_{kl}+{}^{(3)}R-
2m^2(\tr(\sqrt{\hg^{-1}f}-3))] \ ,
\end{eqnarray}
where the square root of the matrix is
defined such that
\begin{equation}
\left(\sqrt{\hg^{-1}f}\sqrt{\hg^{-1}f}\right)^\mu_\nu
=\hg^{\mu\lambda}f_{\lambda \nu} \ .
\end{equation}
Following \cite{Hassan:2011hr} we
perform redefinition of the shift
function \cite{Hassan:2011hr}
\begin{equation}\label{Nidef}
N^i=(\delta^i_j+ND^i_{ \ j})n^j
\end{equation}
for new shift functions $n^i$. Then we
demand that the resulting theory is
linear in $N$.
 In other
words, we demand that
\begin{equation}
N (\sqrt{\hg^{-1} f})^\mu_\nu=
\mathbf{A}_\nu^\mu+ N
\mathbf{B}_\nu^\mu \ ,
\end{equation}
where the matrices $\bA,\bB$ do not
depend on $N$ while it can depend on
$n^i$. Then
\begin{equation}\label{hgeta}
(\hg^{-1}\eta) =\frac{1}{N^2}
\bA^2+\frac{1}{N}(\bA \bB+\bB \bA)+
\bB^2 \ .
\end{equation}
 Following \cite{Hassan:2011hr} we introduce
the matrix  notation, where $n$ denotes
column vector and $n^T$ its transpose.
Further, $\eta=diag (-1,\mathbf{I})$
and $\bI_{ij}=\delta_{ij} \ ,
\bI^{-1}_{ij}=\delta^{ij}$.
Introducing (\ref{Nidef})  into
(\ref{hgeta})   and comparing we find
\cite{Hassan:2011hr}
\begin{eqnarray}\label{bAB}
\bA&=&\frac{1}{\sqrt{1-n^T\bI n}}
\left(\begin{array}{cc} 1 & n^T \bI \\
-n & -n n^T \bI \\ \end{array}\right) \
,
\nonumber \\
\bB&=&\left(\begin{array}{cc} 0 & 0 \\
0 & \sqrt{(g^{-1}-Dn n^T D^T)\bI} \\
\end{array}\right) \ .  \nonumber \\
\end{eqnarray}
Inserting (\ref{Nidef}) and (\ref{bAB})
into (\ref{nonmassact})  we obtain
following action
\begin{eqnarray}\label{mass2red}
S&=&M_p^2\int d^4x \sqrt{g}N
[\tilde{K}_{ij}\mG^{ijkl}\tilde{K}_{kl}
+\bD_{ij}\mG^{ijkl}K_{kl}+
K_{ij}\mG^{ijkl}\bD_{kl}
+\bD_{ij}\mG^{ijkl}\bD_{kl}+\nonumber \\
&+&{}^{(3)}R - 2m^2(
 \sqrt{1-n^i\delta_{ij}n^j}+N
\tr \sqrt{g^{-1}\bI-Dnn^T D^T\bI} -3)
]  \ , \nonumber \\
\end{eqnarray}
where
\begin{eqnarray}
\bD_{ij}&=&-\frac{1}{2N}(\nabla_i (N
g_{jk}D^k_{ \ l} n^l)+ \nabla_j (N
g_{ik}D^k_{ \ l} n^l))=\bD_{ji} \ ,
\nonumber
\\
\tilde{K}_{ij}&=&\frac{1}{2N}
(\partial_t g_{ij}-\nabla_i
n_j-\nabla_j n_i) \  ,
\nonumber \\
\end{eqnarray}
and where we used the fact that
\begin{equation}
\bA^\mu_\mu= \sqrt{1-n^i\delta_{ij}n^j}
\ , \quad \bB^\mu_\mu= \tr
\sqrt{g^{-1}\bI-Dnn^T D^T\bI} \ .
\end{equation}
 Our goal is to
perform the Hamiltonian analysis of the
theory defined by the action
(\ref{mass2red}).
 As follows
from (\ref{mass2red}) the momenta
conjugate to $g_{ij}$ take the form
\begin{eqnarray}
\pi^{ij}&=&M_p^2\sqrt{g}\mG^{ijkl}(\tilde{K}_{kl}+\bD_{kl})
\ .
\nonumber \\
\end{eqnarray}
Then it is easy to find the Hamiltonian
in the form
\begin{eqnarray}
H&=&\int d^3\bx
\left(N\mH_T+\mH_j(\delta^j_i+N D^j_{ \
i})n^i +2
M_p^2m^2\sqrt{g}\sqrt{1-n^i\delta_{ij}n^j}\right)
\ ,  \nonumber \\
\mH_T&=& \frac{1}{M_p^2\sqrt{g}}
\pi^{ij}\mG_{ijkl}\pi^{kl}
-M_p^2\sqrt{g} {}^{(3)}R+2M_p^2
m^2\sqrt{g}(\tr \sqrt{(g^{-1}\bI- Dn
n^TD^T\bI)}-3) \ .
 \nonumber \\
\end{eqnarray}
Due to the absence of the time
derivatives of $N,n^i$ in the action
(\ref{mass2red}) we see that this
theory possesses following collection
of the  primary constraints
\begin{equation}\label{primcon}
\pi_N\approx 0 \ , \quad  \pi_i\approx
0 \  .
\end{equation}
Now the crucial point is  the
requirement of the preservation of
these primary constraints
(\ref{primcon})
 during the
time evolution of the system.
Explicitly, the requirement of the
preservation of the constraints
$\pi_N\approx 0,\pi_i\approx 0$ implies
\begin{eqnarray}\label{secpiN}
\partial_t\pi_N&=&\pb{\pi_N, H}=-\mH_T-\mH_j D^j_{ \ i} n^i\equiv
 -\bar{\mH}_T
\approx 0 \ ,
\nonumber \\
\partial_t\pi_i&=&\pb{\pi_i,H}=\left(-\mH_j +2M_p^2 m^2\sqrt{g}
\frac{\delta_{jk}n^k}{\sqrt{1-n^i\delta_{ij}n^j}}\right)
\left(\delta^j_i+N \frac{\partial
(D^j_{  \ k} n^k)}{\partial n^i}\right)
\ ,
 \nonumber \\
\end{eqnarray}
where we used the fact that
\begin{eqnarray}
\frac{\delta \bA^\mu_\mu}{\delta n^i}=
-\frac{\delta_{ij}n^j}{\sqrt{1-n^i\delta_{ij}n^j}}
\ ,  \quad
\frac{\delta
\bB^\mu_\mu}{\delta n^i}=
-\frac{n^k \delta_{kn}}{\sqrt{1-n^i\delta_{ij}n^j}}
\frac{\partial}{\partial n^i}(D^n_{ \
p} n^p)
\nonumber \\
\end{eqnarray}
and   also the fact that $D$ obeys the
equation \cite{Hassan:2011hr}
\begin{equation}
\sqrt{1-n^T\bI n}D= \sqrt{(g^{-1}-Dn
n^T D^T)\bI} \ .
\end{equation}
Since the expression $\delta^j_i+N
\frac{\partial (D^j_l n^l)}{\partial
n^i}$ is the Jacobian of the
transformation (\ref{Nidef}) it is
non-zero. Then we find that it is
natural to introduce following
 secondary constraint
\begin{eqnarray}
\tilde{\mH}_i= \mH_i- 2M_p^2
m^2\sqrt{g}
\frac{\delta_{ij}n^j}{\sqrt{1-n^i\delta_{ij}n^j}}
\ .
 \nonumber \\
\end{eqnarray}
However using this constraint we observe
that the constraint $\bar{\mH}_T$ can
be written as
\begin{equation}
\bar{\mH}_T= \mH_T+2m^2
M_p^2\sqrt{g}\frac{n^i\delta_{ij}D^j_{
\ k}n^k}{\sqrt{1-n^i
\delta_{ij}n^j}}+\tmH_i D^i_{ \ j}n^j
\end{equation}
and we see that it is natural to
introduce new independent constraint
$\tmH_T$ defined as
\begin{equation}
\tmH_T=
\mH_T+2m^2 M_p^2\sqrt{g}\frac{n^i\delta_{ij}D^j_{ \ k}n^k}{\sqrt{1-n^i
\delta_{ij}n^j}} \ .
\end{equation}
The reason why we consider $\tmH_T$
instead $\bar{\mH}_T$ is that
  the expression
$\mH_i$ is included in  the constraint
$\bar{\mH}_T$  which  makes the
calculation of the Poisson brackets
between the constraints $ \bar{\mH}_T$
rather awkward.

Collecting all these results together we find
 the total
Hamiltonian in the form
\begin{eqnarray}\label{Htotal}
H_T&=&\int d^3\bx (N\mH_T+
\mH_j(\delta^j_i+D^j_{ \ i})n^i + 2
M_p^2m^2\sqrt{g}\sqrt{1-n^i\delta_{ij}n^j}
+ \nonumber
\\
&+& v_N\pi_N+ v^i\pi_i+
u^T\tilde{\mH}_T+ u^i
\tilde{\mH}_i) \ , \nonumber \\
\end{eqnarray}
where $v_N,v^i,u^T,u^i$ are Lagrange
multipliers corresponding to the
collection of all constraints
$\pi_N,\pi_i,\tilde{\mH}_T,\tilde{\mH}_i$.
We see  that $N$ appears linearly in
the total Hamiltonian (\ref{Htotal}).
Finally note that we can express the
total Hamiltonian (\ref{Htotal}) using
the constraints  $\tmH_T,\tmH_i$ as
\begin{eqnarray}
H_T&=&
\int d^3\bx
\left[\tilde{u}^T\tmH_T+\tilde{u}^i
\tmH_i+
 +
 2M_p^2m^2\sqrt{g}\frac{1}{\sqrt{1-n^i\delta_{ij}n^j}}
+ v_N\pi_N+ v^i\pi_i\right]\equiv   \nonumber \\
&\equiv & \int d^3\bx
\left[\mH_0+\tilde{u}^T\tmH_T+\tilde{u}^i
\tmH_i+
  v_N\pi_N+ v^i\pi_i\right] \ ,
\end{eqnarray}
where we defined shifted Lagrange
multipliers
\begin{equation}
\tilde{u}^T=u^T+N \ , \quad
\tilde{u}^i=n^i+D^i_{\ j}n^j+u^i \
\end{equation}
and the bare Hamiltonian $H_0$ as
\begin{equation}
H_0= 2M_p^2m^2\int d^3\bx
\sqrt{g}\frac{1}{\sqrt{1-n^i\delta_{ij}n^j}}
\ .
\end{equation}
 To proceed
further we have to check the stability
of all constraints. To do this we need
following Poisson brackets
\begin{eqnarray}
\pb{\pi_N,\tilde{\mH}_T}&=&0 \ , \quad
\pb{\pi_N,\tilde{\mH}_i}=0 \ ,
\nonumber \\
\pb{\pi_i,\tilde{\mH}_T}&=&
-2m^2
M_p^2\sqrt{g}\frac{\delta_{ij}D^j_{\
n}n^n}{\sqrt{1-n^i\delta_{ij}n^j}}
-
 2m^2 M^2_p \delta_{ij}n^j
\frac{n^k\delta_{kl}D^l_{ \ m}n^m}
{(1-n^i\delta_{ij}n^j)^{3/2}}\equiv
\triangle_{\pi_i\tmH_T} \ , \nonumber
\\
\pb{\pi_i,\tilde{\mH}_j}&=& 2M_p^2 m^2
\sqrt{g}\left(\frac{\delta_{ij}}{\sqrt{1-n^i\delta_{ij}
n^j}}+\frac{\delta_{ik}n^k\delta_{il}n^l}
{(1-n^i\delta_{ij}n^j)^{3/2}}\right)\equiv
\triangle_{\pi_i \tmH_j} \ .
\nonumber \\
\end{eqnarray}
Let us comment  these results. First of
all we see that the Poisson bracket
between $\pi_i$ and $\tmH_j$ is
non-zero on the whole phase space which
implies that $\pi_i$ and $\tmH_j$ are
the second class constraints.  The
situation is more complicated in case
of the Poisson bracket between $\pi_i$
and $\tmH_T$ since this Poisson bracket
vanishes on the subspace $n^i=0$.
However this is the isolated point of
the measure zero so that we can again
say that at the generic point of the
phase space $\pi_i$ and $\tmH_T$ are
the second class constraints.

For further analysis it is convenient
to introduce the smeared form of the
constraints $\tmH_T,\tmH_i$
\begin{equation}
\bT_T(X)=\int d^3\bx X(\bx) \tmH_T(\bx)
\ , \quad \bT_S(X^i)= \int d^3\bx
X^i(\bx)\tmH_i(\bx) \ ,
\end{equation}
where $X,X^i$ are test functions.
 Note that in the  case of the
general relativity we have the
constraints
\begin{equation}
\mH^{GR}_T= \frac{1}{M_p^2\sqrt{g}}
\pi^{ij}\mG_{ijkl}\pi^{kl}
-M_p^2\sqrt{g} {}^{(3)}R \ , \quad
\mH_i^{GR}=-g_{ij}\nabla_k\pi^{jk} \
\end{equation}
whose smeared forms have following
algebra of the  Poisson brackets
\begin{eqnarray}
\pb{\bT_T^{GR}(X),\bT_T^{GR}(Y)}&=&
\bT_S^{GR}((X\partial_j Y-Y\partial_j
X)g^{ji}) \ ,\nonumber \\
\pb{\bT_S^{GR}(X^i),\bT_T^{GR}(Y)}&=&
\bT_T^{GR}(X^i\partial_i Y) \ ,
\nonumber \\
\pb{\bT_S^{GR}(X^i),\bT_S^{GR}(Y^j)}&=&
\bT_S^{GR}(X^i\partial_i
Y^j-X^i\partial_i Y^j) \ . \nonumber \\
\end{eqnarray}
It is important to stress that the
right sides of these Poisson brackets
are proportional to the constraints and
consequently they vanish on the
constraints surface. In other words,
the constraints in the general
relativity are the first class
constrains which is the manifestation
of the fact that general relativity is
the completely constrained system.

Returning to the case of the non-linear
 massive gravity  we now
determine the Poisson brackets between
the constraints $\bT_T(X),\bT_S(X^i)$.
Firstly we obtain
\begin{eqnarray}
\pb{\bT_T(X),\bT_T(Y)}
&=&\bT_S((X\partial_j Y-Y\partial_j
X)g^{ji}) +
\nonumber \\
&+& 2m^2 M_p^2\int d^3\bx (X\partial_i
Y-Y\partial_i X)g^{ij}
\frac{\sqrt{g}\delta_{jk}n^k}{\sqrt{1-n^i\delta_{ij}n^j}}
\equiv \triangle_{TT}(N,M)
\ . \nonumber \\
\end{eqnarray}
Then we calculate following Poisson
bracket
\begin{eqnarray}
\pb{\bT_S(X^i),\bT_S(Y^j)}&=&
\bT_S(X^i\partial_i Y^j-Y^i\partial_i
X^j)+\nonumber
\\
&+& \int d^3\bx (X^i\partial_i
Y^j-Y^i\partial_i X^j)\frac{2M_p^2
m^2\sqrt{g}\delta_{jk}n^k}{\sqrt{1-n^i
\delta_{ij}n^j}}-\nonumber \\
&-& \int d^3\bx
\partial_k\left[\frac{2M_p^2
m^2\delta_{ij}n^j}{\sqrt{1-n^i\delta_{ij}n^j}}\right]
(X^k Y^i-X^i Y^k) \nonumber \\
&\equiv & \triangle_{SS}(N^i,M^j) \ .
 \nonumber \\
\end{eqnarray}
In the same way we determine the
Poisson bracket
\begin{eqnarray}
\pb{\bT_S(X^i),\bT_T(Y)}=\bT_T(X^i\partial_i
Y)+\Phi_{ST}(n^i,X^i,Y)\equiv
\triangle_{ST}(X^i,Y) \ ,  \nonumber \\
\end{eqnarray}
where the functional
$\Phi_{ST}(n^i,g,N^i,M)$ depends on
$n^i,g_{ij}$.
 Finally we calculate following  Poisson brackets
\begin{eqnarray}
\pb{\bT_T(X),H_0}&=&-8m^2 \int d^3\bx
X\frac{\pi^{ij}g_{ji}}
{\sqrt{1-n^i\delta_{ij}n^j}}\equiv
\triangle_{TH}(X)\neq
0 \ , \nonumber \\
\pb{\bT_S(X^i),H_0}&=&2m^2 M_p^2 \int
d^3\bx X^k
\partial_k\left(\frac{1}{\sqrt{1-n^i\delta_{ij}n^j}}
\right)\sqrt{g}\equiv
\triangle_{SH}(X^i)\neq 0 \
 \nonumber \\
\end{eqnarray}
which are non-zero  on the whole phase
space.

 Now
we are ready to analyze the time
evolution of the constraints
$\pi_N,\pi_i$
\begin{eqnarray}\label{timepi}
\partial_t \pi_N&=&\pb{\pi_N,H_T}\approx 0 \ ,
\nonumber \\
\partial_t\pi_i&=&\pb{\pi_i,H_T}\approx
\int d^3\bx
(u^T\triangle_{\pi_i,\tmH_T}+
u^j\triangle_{\pi_i,\tmH_j})=0 \ .
\nonumber \\
\end{eqnarray}
From the first equation we see that
$\pi_N$ is the first class constraint
while the second equation shows that
$\pi_i$ is the second class constraint.
 On the other
hand the time evolution of the
constraint $\bT_T(X),\bT_S(X^i)$
implies
\begin{eqnarray}\label{timeb}
\partial_t\bT_T(X)&=&
\pb{\bT_T(X),H_T}=\triangle_{TH}(X)+
\triangle_{TT}(N,u^T)+\nonumber \\
&+& \triangle_{TS}(X,u^i)+
\triangle_{T,\pi_i}(X,u^i)=0 \ ,
\nonumber
\\
\partial_t\bT_S(X^i)&=&
\pb{\bT_S(X^i),H_T}=
\triangle_{SH}(x^i)+\triangle_{ST}(X,u^i)+
\nonumber \\
&+& \triangle_{SS}(X^i,u^j)+
\triangle_{S\pi_i}(X^i,v^j)=0 \ .
 \nonumber \\
\end{eqnarray}
For generic situation when $n^i\neq 0$
we have $7$ equations (\ref{timepi})
and (\ref{timeb})
  for
unknown $7$ Lagrange multipliers
$u^T,u^i,\pi^i$. In other words we have
$7$ second class constraints
$\tmH_T,\tmH_i,\pi_i$ while we have one
the first class constraint $\pi_N$. The
constraints $\tmH_i=0 , \pi_i=0$ allow
to eliminate $n^i,\pi_i$ in terms of
the phase space variables
$g_{ij},\pi^{ij}$. The constraint
$\pi_N\approx 0$ can be gauge fixed
with the condition $N=0$ and hence
$N,\pi_N$ are eliminated as well.
Finally the second class constraint
$\tmH_T=0$ eliminates  one phase space
degree of freedom so that we have $11$
physical degrees of freedom. Since the
massive gravity should have $10$
physical degrees of freedom we see that
there is one  extra $1/2$ degree of
freedom whose physical interpretation
 is unclear. We  would like to
stress  that there is another example
of the system with single second class
constraint per space-time point which
is the chiral boson
\cite{Floreanini:1987as,Henneaux:1988gg}.
The existence of the single second
class constraint in given system will
be shown in the appendix.
\section{Conclusion}\label{fourth}
In this section we  outline our
results. We developed the Hamiltonian
formalism for non-linear massive
gravity in the formulation presented in
\cite{Hassan:2011hr}. We made an
emphasis on the careful analysis of the
preservation of the constraints during
the time evolution of the system. We
showed that the Hamiltonian constraint
is the second class constraint and
hence its time evolution does not
generate an additional constraint. As a
result this theory possesses the number
of degrees of freedom corresponding to
the massive gravity together  with
 one extra $1/2$ mode whose physical
 origin is unclear. In other words we
 mean that even if the proposal
 suggested in \cite{Hassan:2011hr} is
 very promising it is not sufficient
 for the complete elimination of all
non physical degrees of freedom.

We should also make  comments about the
relation of our work to the paper
\cite{Hassan:2011hr}. The authors claim
that after performing the redefinition
of the shift function it is possible to
integrate out these shift functions so
that we derive the  massive gravity
action that is function of the physical
degrees of freedom only and which is
linear in $N$. Then clearly the
requirement of the preservation of the
primary constraint $\pi_N\approx 0$
generates the secondary constraint
which we denote as $\Phi$. However the
crucial point is that the Poisson
bracket $\pb{\Phi(\bx),\Phi(\by)}$
cannot be zero or proportional to
$\Phi(\bx)$. In fact, since $\Phi$
contains the standard general
relativity Hamiltonian constraint
together with additional terms we
expect that the Poisson brackets
$\pb{\Phi(\bx),\Phi(\by)}$ is
proportional to $ -\nabla_i \pi^{ij}$
and to some additional terms. In case
of the general relativity the
expression $ -\nabla_i \pi^{ij}$ is
proportional to the generator of the
spatial diffeomorphism which is the
constraint as well and hence the
algebra of constraints in general
relativity is closed. In other words,
they are the first  class constraints.
However in case of the Hamiltonian
found \cite{Hassan:2011hr} there are no
such constraints since the shift
functions have been integrated out. As
a result we mean that it is appropriate
to interpret $\Phi$ as the second class
constraint with all physical
consequences.

As the possible extension of our work
we mean that it would be certainly very
interesting to perform the Hamiltonian
analysis of the non-linear massive
gravity action written with the help of
the St\"{u}ckelberg fields
\cite{deRham:2011rn}. We hope to return
to this problem in near future, at
least in case of the $1+1$ dimensional
toy model of the massive gravity action
proposed recently in
\cite{deRham:2011rn}.

\begin{appendix}
\section{Appendix:
\\
Hamiltonian Analysis of Chiral
Boson}\label{appendix}
 In
this appendix we would like to give an
example of the system with single
second class constraint. Let us
consider the Lagrangian for the scalar
field in two dimensions
\begin{equation}
\mL=\frac{1}{2}(\partial_\tau\phi)^2-
\frac{1}{2}(\partial_\sigma\phi)^2 \ .
\end{equation}
It is easy to find corresponding
Hamiltonian
\begin{equation}
H=\frac{1}{2}p_\phi^2+(\partial_\sigma
\phi)^2 \ .
\end{equation}
Now we subject this theory with the
chirality constraint
\begin{equation}
\mathcal{C}=p_\phi-\partial_\sigma\phi=0
\ .
\end{equation}
It turns out that this is the second
class constraint since
\begin{equation}
\pb{\mathcal{C}(\sigma),\mathcal{C}(\sigma')}=
-2\partial_\sigma
\delta(\sigma-\sigma') \ .
\end{equation}
Clearly the complete Hamiltonian
treatment of given theory consists in
the replacement of the Poisson brackets
with corresponding Dirac brackets.
However the goal of this  appendix was
to give an explicit example of the well
known physical system with the single
second class constraint.

\end{appendix}

\

 \noindent {\bf
Acknowledgements:}
 This work   was
supported by the Czech Ministry of
Education under Contract No. MSM
0021622409. \vskip 5mm

\end{document}